\documentclass[12pt]{amsart}
\usepackage{amsmath}
\usepackage{amsxtra}
\usepackage{amscd}
\usepackage{amsthm}
\usepackage{amsfonts}
\usepackage{amssymb}
\usepackage{eucal}
\usepackage{epsfig}
\usepackage{graphics}
\usepackage[dvips]{color}
\usepackage{epsf,graphics,mathrsfs,yfonts,eufrak,simplewick}
\usepackage{pdfpages}
\usepackage{psfrag}
\usepackage[OT2,T1]{fontenc}

\def\'{\char126}
\def\`{\char127}

\def\({\left(}
\def\){\right)}

\newcommand{\cb}{\mathbf{c}}
\newcommand{\bb}{\mathbf{b}}

\newcommand{\tb}{\mathbf{t}}

\newcommand{\nn}{\nonumber}

\newcommand{\R}{{\mathbb R}}

\newcommand{\C}{{\mathbb C}}

\newcommand{\slt}{\mathfrak{sl}_2}
\newcommand{\slth}{\widehat{\mathfrak{sl}}_2}

\newcommand{\Tr}{{\rm Tr}}

\newcommand{\Ker}{\mathop{\rm Ker}}

\newenvironment{tenumerate}{
  \begin{enumerate}
  
  }{\end{enumerate}}
\newcommand{\bi}{\begin{tenumerate}}
\newcommand{\ei}{\end{tenumerate}}
\newcommand{\isoto}[1][]%
{{\mathop{\buildrel{\sim}\over\longrightarrow}\limits_{#1}}}

\def\[{\left[}
\def\]{\right]}
\newcommand{\la}{\lambda}

\newcommand{\al}{\alpha}

\numberwithin{equation}{section}

\def\half{\textstyle{\frac  1 2}}

\def\bn{\mathbf{n}}

\def\bi{\mathbf{i}}

\usepackage[dvipsnames]{xcolor}
 \usepackage[enableskew]{youngtab}
\begin{document}
\begin{title}[OPE for XXX]{
OPE for XXX}

\end{title}
\date{today}
\author{Philippe Di Francesco and Fedor Smirnov
}
\address{PhDF: IPhT, Univ. Paris Saclay, CNRS, CEA, \\F-91191 Gif-sur-Yvette, France\\  
and Dept of Mathematics, Univ. of Illinois MC-382, \\ Urbana, IL 61821, U.S.A.}
\email{philippe.difrancesco@gmail.com}
\address{FS\footnote{Membre du CNRS} Sorbonne Universit\'e, UPMC Univ. Paris 06\\ CNRS, UMR 7589, LPTHE\\F-75005, Paris, France}
\email{smirnov@lpthe.jussieu.fr}

\dedicatory{ In memory of Ludwig Dmitrievich Faddeev}
\begin{abstract}
We explain a new method for finding the correlation functions for the XXX model
which is based on the concepts of Operator Product Expansion of Quantum Field Theory on one hand 
and of fermionic bases for the XXX spin chain on the other. 
With this method we are able to perform computations for up to 11 lattice sites. 
We show that these ``experimental" data allow to guess exact formulae
for the OPE coefficients.
\end{abstract}

\maketitle

\section{Introduction}

This paper deals with the isotropic Heisenberg spin chain. 
However, even dealing with a lattice model, it is important to have in mind
certain general ideas coming from Quantum Field Theory (QFT). One of them is the 
operator product expansion (OPE). 

Suppose for a given QFT we know a complete set of local operators $\mathcal{O}_i(x)$.
Then for small $x$ we have an expansion
\begin{align}
\mathcal{O}_i(x)\mathcal{O}_j(0)=\sum\limits_kC_{i,j}^k(x)\mathcal{O}_k(0)\,.\nn
\end{align}
We take the OPE quite symbolically leaving aside the necessity of time ordering, the convergence issues
{\it etc}. An important feature of the OPE  is that it is defined
only by the short distance, ultra-violet (UV) nature of the theory. The infra-red (IR) environment becomes relevant   
when the correlation functions are computed. In the OPE approach, it enters 
through the one-point functions only:
\begin{align}\langle  \mathcal{O}_i(x)\mathcal{O}_j(0)\rangle_\mathrm{env}=\sum\limits_kC_{i,j}^k(x)\langle\mathcal{O}_k(0)\rangle_\mathrm{env}\,.
\label{OPE}
\end{align}

A first serious problem concerning the application of OPE consists in finding a convenient way to enumerate the
local operators. It is well known that in a two-dimensional conformal field theory (CFT) they correspond to representations
of the Virasoro algebra. A less known example is provided by the sine(h)-Gordon model for which there exists a fermionic 
basis for the local operators where all the one-point functions are expressed as determinants
\cite{HGSV,NS}.
In the UV limit the relation of the fermionic basis to the Virasoro basis is known, and this allows in principle to
apply Perturbed CFT methods to the computation of the two-point functions.

On the other hand, let us imagine a situation when for a large number of IR environments the one-point functions
and two-point functions are known. Then the relations \eqref{OPE} provide equations for the coefficients of the OPE.
One can dream that these equations could be sufficient to entirely fix all the coefficients of the OPE. For the moment
this program does not look realistic for the QFT. However, we shall show that it is applicable to integrable
lattice models.

Generally, the OPE does not depend on the IR environment. However, one can consider  classes
of IR environments possessing certain symmetries. For example, they can respect translational invariance.
In this case the spatial derivatives of one-point functions vanish, and we can drop them in the right hand 
side of the OPE. We shall deal with much wider symmetries of this kind, and write the OPE modulo 
symmetries as 
\begin{align}
\mathcal{O}_i(x)\mathcal{O}_j(0)\equiv\sum\limits_kC_{i,j}^k(x)\mathcal{O}_k(0)\,.\nn
\end{align}
Certainly, the symmetries in question will have to be carefully implemented. They can be not only continuous,
but also discrete ones, like $C$-symmetry for instance. 

In this paper we shall consider the celebrated isotropic Heisenberg magnet (XXX spin chain) with the
Hamiltonian
$$H=\sum_{j=-\infty}^{\infty} h_{j,j+1},\quad h_{j,j+1}=\half(\sigma_j^a\sigma_{j+1}^a-1)\,,$$
with usual notations for Pauli matrices $\sigma_j^a$ at site $j$, and where summation over repeated indices is always implied. 
For some technical
reasons which will be explained later we consider the two-point functions of $\slt$-invariant operators. 
The examples will be $\sigma_1^a\sigma_N^a$ and
$h_{1,2}h_{N-1,N}$. 

We consider the correlation functions with arbitrary Matsubara data which will be introduced in the next
section. This is a quite general situation which includes important physical applications: zero-temperature anti-ferromagnet,
(anti)-ferromagnet with non-zero temperature and magnetic field or even the case of generalised
Gibbs ensemble. All these cases correspond to choosing certain special Matsubara data and taking
limits for them.

The set of local operators
in the lattice analogue of OPE is expressed via the fermionic basis. 
Crucial to our study are the cases with simple Matsubara data. 
For these the one-point functions are computed as determinants and there is an efficient procedure
for computing the two-point functions using the quantum inverse scattering method (QISM) techniques
\cite{FST,BIK,Slavnov}. This allows to obtain
an over-determined system of equations for the OPE coefficients. We check that these systems have
solutions up to $N=11$ and compute all the OPE coefficients. These coefficients can now be used for any
Mastubara data. 

It will be very interesting to guess general formulae for the OPE coefficients from our ``experimental"
data. This part of the work is still under way, but we shall present here some preliminary results.

The paper is organised as follows. In Section~2 we formulate the problem of computing correlation functions on
a cylinder (Matsubara environment). In Section~3 the fermionic basis and its restriction to our problem
are explained. In Section~4 we present an efficient computation procedure for small Matsubara chains.
In Section~5 we further restrict the fermionic basis. The results of computations and formulae for
some elements of OPE are given in Section~6.

\section{Formulation of the problem}

Consider the following square lattice wrapped onto a cylinder $\R\times S^1$
\begin{figure}
\centerline{\includegraphics[width=10cm]{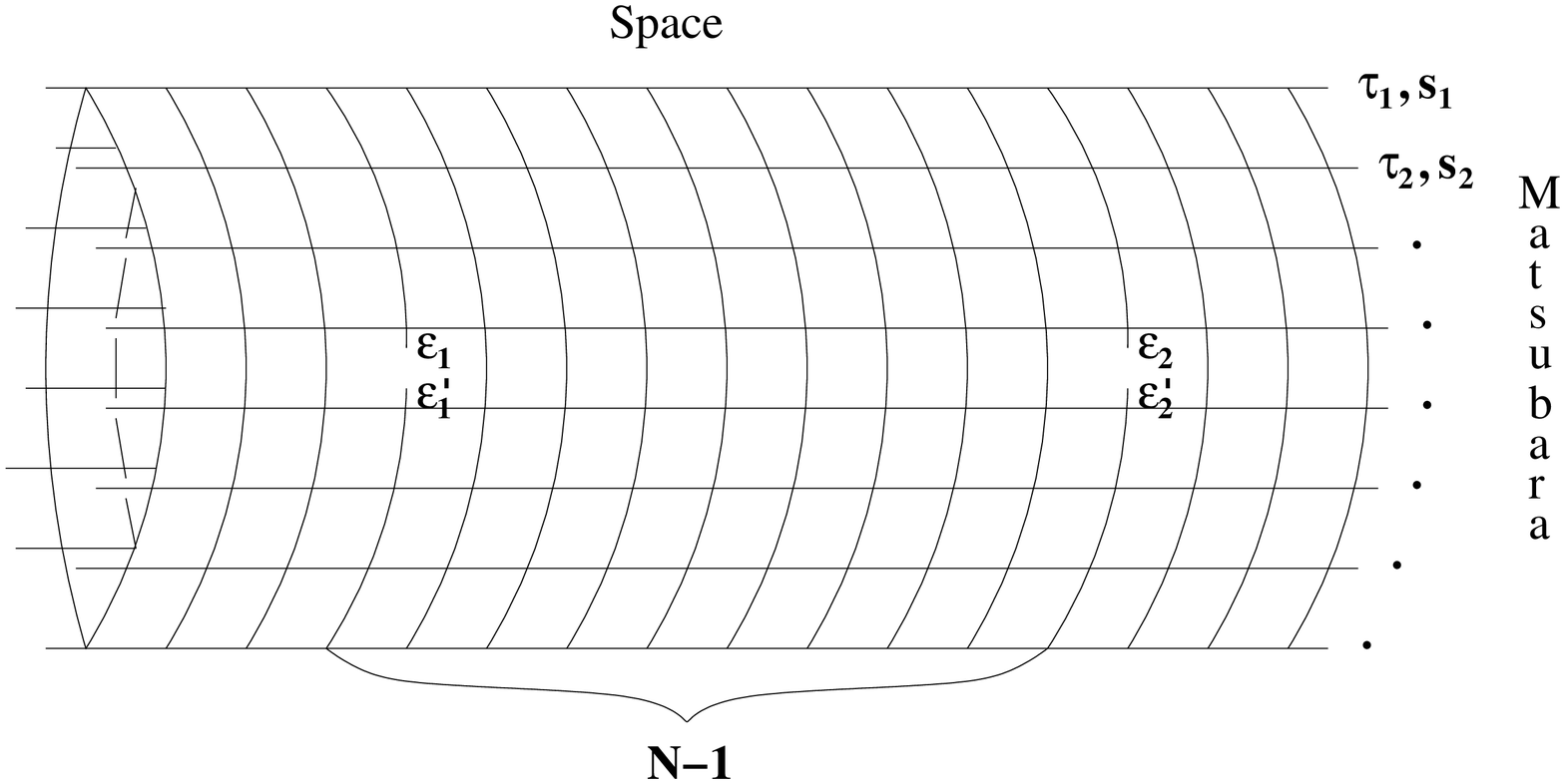}}
\caption{\ }
\end{figure}

We associate the space $\mathbb{C}^2$ to each horizontal lattice site (along each of the parallel circles), and
the space $\mathbb{C}^{2s_\mathbf{m}+1}$ together with the complex
parameter $\tau_\mathbf{m}$ to the $\mathbf{m}$-th site
of the compact Matsubara lattice (along horizontal lines). 
Denote by $\pi_{\la}^{2s}$ the evaluation representation of the Yangian 
with the dimension $2s+1$ and evaluation parameter $\la$.
Every crossing corresponds to the $\slt$ Yangian $R$-matrix in 
the tensor product of two evaluation representations: $\pi_0^1$ and $\pi_{\tau_\mathbf{m}-1/2}^{2s_\mathbf{m}}$. Summation is
implied over all indices except the four fixed: $\epsilon_1,\epsilon_2,\epsilon_1', \epsilon_2'$. 
We denote the partition function defined by the picture of Fig.1 by $Z_{\epsilon_1,\epsilon_2}^{\epsilon_1',\epsilon_2'}$, then
the expectation value of  $\sigma_1^a\sigma_N^a$ (correlation function of spins) with the Matsubara data $\{s_\mathbf{m},\tau_\mathbf{m}\}$ is defined by
$$G^{\sigma\sigma}(N)=1+\frac 2 Z\(Z_{+-}^{-+}+Z_{-+}^{+-}-2Z_{+-}^{+-}\)\,,$$
where $Z$ is the usual partition function (without insertions). 
A similar formula based on the partition function with four operator insertions (at the sites $1,2,N-1,N$) is easy to write for
the expectation value $G^{hh}(N)$  of $h_{1,2}h_{N-1,N}$ (correlation function of densities of the Hamiltonian).
The Matsubara direction has finite size $\bn$ while the horizontal (space) direction is infinite. 
The latter infinity may cause trouble, so we regularise the horizontal space by  assuming it has finite length equal to $2L$ 
and imposing periodic boundary conditions (this amounts to putting our lattice on a torus),
and then by finally sending $L\to\infty$. 

Let us be more formal. Consider the representation of Yangian $\pi _\mathbf{S}$ which is the tensor product of
$2L$ representations $\pi^1_0$. Consider in
addition the representation $\pi _\mathbf{M}$ which is the tensor product of the $\bn$ representations
$\pi^{2s_\mathbf{m}}_{\tau_\mathbf{m}-1/2}$, $\mathbf{m}=1,2,...,\bn$. Define
$$T_{\mathbf{S},\mathbf{M}}=(\pi _\mathbf{S}\otimes\pi _\mathbf{M})(\mathcal{R})\,,$$
with $\mathcal{R}$ being the universal $R$-matrix. Then for any $\mathcal{O}$ acting on a small (compared to $L$)
number of tensor components of the representation space of $\pi _\mathbf{S}$ define
$$Z(\mathcal{O})=\lim_{L\to\infty}\frac{\Tr_{\mathbf{M}}\Tr_{\mathbf{S}}\(T_{\mathbf{S},\mathbf{M}}\mathcal{O}\)}
{\Tr_{\mathbf{M}}\Tr_{\mathbf{S}}\(T_{\mathbf{S},\mathbf{M}}\)}\,.$$
This is the more general definition of the functional $Z$ which was introduced earlier in a particular case.

To the left and to the right of the insertion (see {\it Fig.1}) we have products of Matsubara transfer-matrices
$$T_\mathbf{M}=T_\mathbf{M}(0),\quad T_\mathbf{M}(\la)=\Tr\(\pi_\la^1\otimes \pi_\mathbf{M}(\mathcal{R})\)\,,$$
where the trace is taken with respect to the first tensor component. Generically the transfer-matrix $T_\mathbf{M}$
has a single eigenvector $|\mathrm{max}\rangle$ corresponding to an eigenvalue $T$ of maximal absolute value. 
Suppose the operator $\mathcal{O}$ is localised on the interval $[1,N]$ (like on {\it Fig.1}). 
Clearly, because of the limit $L\to \infty$
$$
Z(\mathcal{O})=\frac{\langle\mathrm{max}|\Tr_{[1,N]}\(T_{[1,N],\mathbf{M}}\mathcal{O}\)|\mathrm{max}\rangle}
{T^N{\langle\mathrm{max}|\mathrm{max}\rangle}}\,.
$$
This quantity was actually computed using the fermionic basis in \cite{HGSIII}. 
The computation is purely algebraic and uses only the fact that $|\mathrm{max}\rangle$ is an eigenvector. 
For that reason the computation extends to more general quantities of the form
\begin{align}
Z(\mathcal{O},\Psi)=\frac{\langle\Psi |\Tr_{[1,N]}\(T_{[1,N],\mathbf{M}}\mathcal{O}\)|\Psi\rangle}
{T^N\langle\Psi|\Psi\rangle}\,,
\end{align}
with an arbitrary eigenstate $|\Psi\rangle$ with eigenvalue $T$  of the transfer-matrix $T_\mathrm{M}$.

\section{Fermionic basis}\label{Basis}

The fermionic basis is defined in \cite{HGSII} for the anisotropic XXZ model with the anisotropy parameter 
$\Delta=\frac 1 2(q+q^{-1})$. Taking the limit $q\to 1$ is not a difficult problem. More importantly,  Ref.~\cite{HGSII} deals with the quasi-local
operators of the form $q^{\al\sum_{j=-\infty}^0\sigma^3_j}\mathcal{O}$. Taking the limit $\al\to 0$ in this expression is non trivial, and
requires some explanation. 

In the XXZ case the basic algebraic tool is the quantum affine algebra $U_q(\slth)$ which replaces the Yangian used for
the XXX case. The main result of \cite{HGSII} is the existence of creation operators $\tb^*_j$ (boson), $\bb^*_j$, $\cb^*_j$
(fermions) for $j=1,2,\cdots$. By acting on the primary field $\Phi_\al=q^{\al\sum_{j=-\infty}^0\sigma^3_j}$ these
operators create a basis in the space of quasi-local operators for which the expectation values of the type
considered in the previous section are easily computed. It should be noticed that 
contrary to the fermions the operators $\tb^*_j$ are of rather simple origin, however they make computations 
much more involved. 

In the limit $\al\to 0$ the operators $\bb^*_j$, $\cb^*_j$ acting on $\Phi_\al$ develop simple poles in $\al$. 
As a consequence, computing expectation values involves using L'Hospital's rules.
This is doable, but leads eventually to the doubling of fermions: one has to consider for every fermion its 
regular part and residue at $\al=0$. However, there is a notable exception of this general picture. The algebra $U_q(\slth)$
contains two finite-dimensional subalgebras isomorphic to $U_q(\slt)$. Consider the operators $\mathcal{O}$
commuting with the action of one of them. In a {\it weak} sense (when inserted in expectation values)
it can be shown that in the limit $\al\to 0$ these operators are created by the action of regular
parts of fermions only. This is a great simplification. Another nice point is that the operators $\tb^*$ do not 
contribute for $\al=0$.

Passing to the XXX case with $\al=0$ we assert that the local operators are created
by $\bb^*_j$, $\cb^*_j$. We use the multi-index notations: for $I=\{i_1,\cdots ,i_n\}$
we define $\bb^*_I=\bb^*_{i_1}\cdots \bb^*_{i_n}$,  $\cb^*_I=\cb^*_{i_n}\cdots \cb^*_{i_1}$. 
We denote by $\#(I)=n$ the cardinality of $I$, and by $|I|=\sum_{\ell=1}^n i_\ell$. Then according
to previous considerations the  invariant operators  are contained in the space $\mathfrak{H}_0$ of operators
with the basis
\begin{align}\bb^*_I\cb^*_J\cdot \mathrm{I}\,,\qquad \#(I)=\#(J)\,,\label{basis}\end{align}
$\mathrm{I}$ standing for the unit operator.
The operator \eqref{basis} is supported on an interval of length at most $|I|+|J|$.

It is important to identify among these the operators which are localised on an interval of length $N$. 
The following first two conditions follow for instance from \cite{completness}:
\begin{align}\#(I)\le [N/2]\,, \quad \mathrm{max}(I\cup J)\le N\,.\label{lessN}
\end{align}
Moreover, we consider $C$-invariant operators (invariant under $\sigma_j^a\to-\sigma_j^a$). For them it can
be shown that the following additional condition can be imposed:
\begin{align}
|I|+|J|\equiv 0\ \ (\mathrm{mod}\ 2)\,.\label{Cinv}
\end{align}
Let us denote the subspace of $\mathfrak{H}_0$  satisfying \eqref{lessN}, \eqref{Cinv} by $\mathfrak{H}^{(N)}_0$.

More sophisticated conditions result from the homogeneous limit of the results of  \cite{FB}.
Consider the  annihilation  operators $\bb_j$, $\cb_j$ canonically conjugated to $\bb^*_j$, $\cb^*_j$, and
annihilating the vacuum.
Introduce the operators
$$Q_m=\sum_{j=1}^{m-1}c_jb_{m-j}\,,\quad m=1,2,\ldots$$
We state that the operators localised on $[1,N]$ are linear combinations of \eqref{basis} satisfying
\begin{align}Q_m\mathcal{O}=0\,,\qquad m>N\,. \label{Qs}\end{align}
Due to the second condition in \eqref{lessN} for given $N$ we can restrict the values of $m$ in the above
to $m\le 2N+1$. 

The main theorem of \cite{HGSIII} states that for $\#(I)=\#(J)$
\begin{align}
Z(\bb^*_I\cb^*_J\cdot \mathrm{I},\Psi)=\det\(\omega_{i_p,j_q}\)_{p,q=1,\ldots, \#(I)}\,,\label{MAIN}
\end{align}
where
$\omega_{i,j}$ is an infinite matrix depending on
$$\mathrm{Matsubara\ data}=\left\{\{s_\mathbf{m},\tau_\mathbf{m}\}_{\mathbf{m}=1}^\bn\,,\ \ |\Psi\rangle\right\}\,.
$$
The exact definition of $\omega_{i,j}$ will be given in the next section.

In our case ($\al=0$) the matrix $\omega_{i,j}$ is symmetric. 
Since we decided to work in a weak sense, this imposes additional 
restrictions. The first of them is obvious: we can consider the space $\mathfrak{H}_0/\iota$, with $\iota$ being the
anti-homomorphism $\bb^*_j\leftrightarrow \cb^*_j$. For example, we can require that $I\le J$ in lexicographical order.
Another consequence is less trivial. The following simple example explains that there are linear relations between
minors of any given symmetric matrix:
\begin{align}
\left|\begin{matrix} \omega_{1,3} & \omega_{1,4}\cr \omega_{2,3} & \omega_{2,4}\end{matrix}\right|-\left|\begin{matrix} 
\omega_{1,2} & \omega_{1,4}\cr \omega_{2,3} & \omega_{3,4}\end{matrix}\right|+\left|\begin{matrix} \omega_{1,2} 
& \omega_{1,3}\cr \omega_{2,4} & \omega_{3,4}\end{matrix}\right|=0\nn\,.
\end{align}
This identity implies that working in a weak sense we can factor out the null-vector
 $$\bigl(b^*_1b^*_2c^*_4c^*_3-b^*_1b^*_3c^*_4c^*_2+b^*_2b^*_3c^*_4c^*_1\bigr)\cdot \mathrm{I}\,.$$
 We look for a general formula for such null-vectors. 
 Consider the operator
 $$C=\sum\limits_{i=1}^\infty c^*_ib_i\,,$$
and the space  $\mathfrak{H}_2$ spanned by $\bb^*_I\cb^*_J\cdot \mathrm{I}$ with $\#(I)=\#(J)+2$.
Then the subspace $\mathfrak{N}=C \mathfrak{H}_2$ of the space $\mathfrak{H}_2$ consists of null-vectors.
Indeed, by taking linear combinations, the formula \eqref{MAIN} can be rewritten as
$$Z(\mathcal{O},\Psi)=
I^*\ e^\Omega \ \mathcal{O}\,,$$
where $\Omega=\sum_{i,j}\omega_{i,j} b_ic_j$ and
the left vacuum $\mathrm{I}^*
$ is annihilated by the right action of creation operators. The fact that $\mathfrak{N}$
consists of null-vectors follows from the commutation of $C$ with $\Omega$.
We do not have a formal proof that belonging to $\mathfrak{N}$ is necessary for a vector to be a null-vector,
but since the computer does not contradict this, we shall assume it is true. 

Let us summarise what we have for the moment. In the weak sense for any local operator $\mathcal{O}$ acting on $N$ sites we have
\begin{align}
\mathcal{O}\equiv 
\sum\limits_\al C_\al v_\al\,,
\label{fermiobasis}
\end{align}
where $v_\al$ is a basis of the space
\begin{align}
\mathfrak{V}^{(N)}=\bigcap\limits_{i=N+1}^{2N+1} \Ker\left(Q_{i}\big\vert_{\mathfrak{H}_0^{(N)}/\iota/\mathfrak{N}}\right)\,.
\label{resspace}
\end{align}
the dimension of this space is much smaller than the dimension of $\mathfrak{H}_0^{(N)}/\iota$. 
We shall see that it is possible to reduce the space further, but with $\mathfrak{V}^{(N)}$ we can already start working.

\section{Explicit computations for small Matsubara chain}

The main goal of this section is to show how to produce a large number of Matsubara data, and subsequently
compute the matrix $(\omega_{i,j})$ and expectation values using them.  

Suppose we are given some Matsubara parameters $\tau_1,s_1,\cdots \tau_\mathbf{n},s_\mathbf{n}$. 
The eigenvectors  $|\Psi\rangle$ are parametrised by Bethe roots $\beta_1,\cdots, \beta_m$. 
They satisfy the Bethe equations which are conveniently written in terms of
the Baxter function $Q(\la)=\prod_{j=1}^m(\la-\beta _j)$:
\begin{align}
a(\beta_j)Q(\beta_j+1)+d(\beta_j)Q(\beta_j-1)=0,\ \ \ j=1,\cdots, m\,,\label{bethe}
\end{align}
where
$$a(\la)=\prod\limits_{\mathbf{m}=1}^\mathbf{n}(\la-\tau_\mathbf{m}-s_\mathbf{m}),\qquad
d(\la)=\prod\limits_{\mathbf{m}=1}^\mathbf{n}(\la-\tau_\mathbf{m}+s_\mathbf{m})
\,.$$
Recall $s_\mathbf{m}$ are given integers, however the subsequent formulae are all analytical in $s_\mathbf{m}$, 
hence we may consider them as arbitrary complex parameters. Then $a(\la), d(\la)$ will be considered as
arbitrary monic polynomials of degree $\mathbf{n}$:
\begin{align}
a(\la)=\la^{\mathbf{n}}+\sum_{j=1}^\mathbf{n}a_j\la^{\mathbf{n}-j},\quad
d(\la)=\la^{\mathbf{n}}+\sum_{j=1}^\mathbf{n}d_j\la^{\mathbf{n}-j}\,.
\end{align}
The usual problem of the theory of quantum integrable models is to find  solutions  to
Bethe equations for the input data
\begin{align}
\mathrm{input}=\{a_1,\cdots a_\mathbf{n},d_1,\cdots d_\mathbf{n}\}\,.
\end{align}
Now solving the Bethe equations we would obtain an equivalent set of parameters to the previously introduced 
\begin{align}\mathrm{Matsubara\ data}=\{\{\tau_\mathbf{m}\}, \{s_\mathbf{m}\},\Psi \}
\equiv\left\{\beta_1,\cdots\beta_m,a_1,\cdots a_\mathbf{n},d_1,\cdots d_\mathbf{n}\right\}\,.\label{data}
\end{align}
For certain well-known reason we set $m\le[\mathbf{n}/2]$. 

The main problem here is that the Bethe equations constitute a complicated system of algebraic equations for 
$\beta_1,\cdots\beta_m$. But we have to realise that our goal is different from the usual one. We do not need to 
find the spectrum for a given chain. Rather we need a stock of solutions of the Bethe equations in order to obtain equations for the 
OPE coefficients. That is why we make the following new choice of input data
\begin{align}
\mathrm{input}=\{\beta_1,\cdots,\beta_m, a_{m+1},\cdots a_\mathbf{n},d_1,\cdots d_\mathbf{n}\}\,.
\end{align}
The number of continuous parameters is the same, but the unknowns are now $a_1,\cdots,a_m$. 
Note that for them the Bethe equations are {\it linear}. 
Solving them is easy, and we obtain as a result some Matsubara data \eqref{data}. 

Given some Matsubara data, we need first of all to construct the infinite matrix $\omega_{i,j}$. 
The latter is conveniently coded by the generating function
\begin{align}
\omega (\la, \mu):=\sum_{i,j=1}^{\infty}\la^{i-1}\mu^{j-1}\omega _{i.j}\,.\label{omcoef}
\end{align}
Let us define the latter in the spirit of \cite{BGKS}. 
Introduce the kernel and ``half-kernel" functions:
$$K(\la)=-\frac 2 {\la^2-1}\,,\qquad H(\la)=\frac 1{ (\la-1)\la}\,.$$
and the measure
$$d m(\la)=\frac {d\la}{1+\mathfrak{a}(\la)}\,,\qquad \mathfrak{a}(\la)=\frac{a(\la)Q(\la+1)}{d(\la)Q(\la-1)}\,.
$$
We need an auxiliary function defined by the integral equation
$$
G(\eta,\mu)=H(\eta-\la)-\frac 1 {2\pi i}\oint _\Gamma K(\eta-\sigma)G(\sigma,\mu)dm(\sigma)\,,
$$
where the contour $\Gamma$ goes around the Bethe roots $\beta_1,\cdots,\beta_m$ and the point $\sigma=\mu$. For a finite 
Matsubara chain we have a finite number of Bethe roots for which the equation above reduces to a linear
system for $G(\beta_j,\mu)$. 
The function $\omega(\la,\mu)$ is given by 
$$\omega (\la, \mu)=\frac 1 {2\pi i}\oint _{\Gamma '}H(\eta-\la)G(\eta,\mu)dm (\eta)-\frac 1 4 K(\la-\mu)\,,$$
with  $\Gamma'$ containing one more point: $\eta=\la$ . It is easy to see that this function is symmetric.
Finally, extracting the coefficients of $\omega(\la,\mu)$ from \eqref{omcoef} allows to compute the right hand side of
\begin{align}\langle\mathcal{O}\rangle_\mathrm{Md}=
\sum\limits_\al C_\al \langle v_\al\rangle_\mathrm{Md}\,,\label{main}
\end{align}
where Md stands for Mastubara data and $\{v_\al\}$ is the reduced fermion basis \eqref{fermiobasis}. 

The left hand side of \eqref{main} is computed by QISM techniques. 
Let us begin by introducing the main character of QISM, the {\it monodromy matrix}. We use the block decomposition 
with respect to the first tensor component
$$(\pi _\la^1\otimes \pi_\mathbf{M})(\mathcal{R})=\begin{pmatrix} A(\la)&B(\la)\\C(\la) &D(\la) \end{pmatrix}\,,$$
with $A(\la),B(\la),C(\la),D(\la)$ acting in the Matsubara space.
We shall denote $A=A(0)$, {\it etc}.
We want to compute
$$\langle\Psi |\Tr_{[1,N]}\(T_{[1,N],\mathbf{M}}\mathcal{O}\)| \Psi\rangle\,,$$
for $\mathcal{O}$ located on sites $[1,N]$. 
For given Matsubara data we write
\begin{align}
&\langle\Psi|=\langle \beta_1,\cdots,\beta_m|=\langle\downarrow|B(\beta_1)\cdots B(\beta_m)
\,,\nn\\
&|\Psi\rangle=| \beta_1,\cdots,\beta_m\rangle=C(\beta_1)\cdots C(\beta_m)|\downarrow\rangle\,.\nn\end{align}
The normalisation is provided by the Gaudin formula below. 
Obviously, for given $\mathcal{O}$ the expression
$\Tr_{[1,N]}\(T_{[1,N],\mathbf{M}}\mathcal{O}\)$ is certain linear combination of monomials of $A,B,C,D$ of
total degree $N$. So, we have to learn how to compute their averages in an efficient way. 

We start with the Slavnov formula for the scalar product of the Bethe covector $\langle \beta_1,\cdots\beta_m|$ with
an off-shell vector
$| \mu_1,\cdots,\mu_m\rangle=C(\mu_1)\cdots C(\mu_m)|\downarrow\rangle$, $\mu_j$ being arbitrary.
The formula is
\begin{align}
 \langle \beta_1,\cdots\beta_m|\mu_1,\cdots, \mu_m\rangle=\prod\limits_{j=1}^md(\beta_j)d(\mu_j)
\frac{\prod\limits_{i,j=1}^m(\beta_j-\mu_i+1)}{\prod\limits_{i<j}(\mu_i-\mu_j)\prod\limits_{i<j}(\beta_j-\beta_i)}\det({\mathcal N})\,,\label{nik}
\end{align}
where the matrix ${\mathcal N}$ has entries
\begin{align}
N_{i,j}=\frac 1{(\mu_i-\beta_j)}\Bigr(
\frac 1{(\mu_i-\beta_j-1)}
-\frac 1{(\mu_i-\beta_j+1)}\mathfrak{a}(\mu_i)
\Bigr)\,.\nn
\end{align}
The left hand side of \eqref{nik} is by definition a symmetric polynomial of $\mu_1,\cdots, \mu_m$, hence the singularities in the right hand side cancel. 
We shall return to this point later. 

The normalisation of the Bethe vectors is given by the Gaudin formula, expressible via L'Hospital rules
from \eqref{nik}. Explicitly, we have
\begin{align}
{\langle \beta_1,\cdots\beta_m|  \beta_1,\cdots\beta_m\rangle}=\prod_{j=1}^ma(\beta_j)
d(\beta_j)\prod_{i\ne j}\frac{\beta_i-\beta_j+1}{\beta_i-\beta_j}\det({\mathcal G})\,,
\end{align}
where the matrix $\mathcal G$ has entries
$$G_{k,l}=\frac{\partial}{\partial\beta_l}\log\mathfrak{a}(\beta_k)\,,\quad k,l=1,\cdots, m\,.$$

We want to compute
$$\langle \beta_1,\cdots\beta_m|X_1X_2\cdots X_N|  \beta_1,\cdots\beta_m\rangle\,,$$
where $X_j$ is one of $A,B,C,D$. We proceed as follows. Start with the Slavnov formula. Use
the following formulae which are easily obtained by methods of QISM. Consider any Bethe covector
$\langle\Phi|$. Introduce the notations $v(\la)=1/\la,\ u(\la)=v(\la)+1$. Provided 
$$f(\mu_1,\cdots, \mu_m)=\langle\Phi|\mu_1,\cdots, \mu_m\rangle\,,$$
is known for arbitrary $\mu_1,\cdots, \mu_m$ we have
\begin{align}
&\langle\Phi|A|\mu_1,\cdots, \mu_m\rangle=\prod_{j=1}^mu(-\mu_j)a(0)f(\mu_1,\cdots, \mu_m)\label{A}\\
&-\sum_{j=1}^kv(-\mu_j)\prod_{r\ne j}^k  u(\mu_j-\mu_r)a(\mu_j)f(\mu_1,\cdots,\widehat{\mu_j},\cdots, \mu_m,0)\,,\nn
\end{align}
\begin{align}
&\langle\Phi|D|\mu_1,\cdots, \mu_m\rangle=\prod_{j=1}^ku(\mu_j)d(0)\prod_{j=1}^kC(\mu_j)|\downarrow\rangle\label{D}\\
&-\sum_{j=1}^mv(\mu_j)\prod_{r\ne j}^m  u(\mu_r-\mu_j)d(\mu_j)f(\mu_1,\cdots,\widehat{\mu_j},\cdots, \mu_m,0)\,,\nn
\end{align}
\begin{align}
&\langle\Phi|C|\mu_1,\cdots, \mu_m\rangle=\Bigl(\sum_{j=1}^kv(-\mu_j)\prod_{r\ne j}^ku(-\mu_r)(\mu_r-\mu_j)a(0)d(\mu_j)\label{C}\\&+
v(\mu_j)\prod_{r\ne j}^ku(\mu_r)(\mu_j-\mu_r)d(0)a(\mu_j)\Bigr)f(\mu_1,\cdots,\widehat{\mu_j},\cdots, \mu_m)\nn\\
&+\sum_{j>i}\Bigl(d(\mu_i)a(\mu_j)v(-\mu_i)v(-\la)u(\mu_j-\mu_i)\prod_{r\ne i,j}^ku(\mu_r-\mu_i)u(\mu_j-\mu_r)\nn\\
&+a(\mu_i)d(\mu_j)v(-\mu_j)v(\mu_i)u(\mu_i-\mu_j)\prod_{r\ne i,j}^ku(\mu_i-\mu_r)u(\mu_r-\mu_j)\Bigr)\nn\\&\times
f(\mu_1,\cdots,\widehat{\mu_i},\cdots,,\cdots,\widehat{\mu_j},\cdots, \mu_m,0)\,.\nn
\end{align}
\begin{align}
&\langle\Phi|B|\mu_1,\cdots, \mu_m\rangle=f(\mu_1,\cdots,\mu_m,0)\,.\qquad\qquad\qquad\qquad\qquad\label{B}
\end{align}
Notice that in the last two formulae the number of arguments changes.
Using these formulae we compute inductively
$$\langle \beta_1,\cdots\beta_m|X_1X_2\cdots X_N|  \mu_1,\cdots\mu_m\rangle\,,$$
and then set $\mu_j=\beta_j$, $j=1,\cdots,m$. 

Our goal is to implement the formulae above on a computer. 
To proceed fast we have to avoid symbolic computations as much as possible. 
The  input data are chosen numerically as a set of random integers from $1$ to $10$. 
Still in the procedure described above we cannot immediately exclude the symbolic variables $\mu_j$. 
It has been said that 
$ \langle \beta_1,\cdots\beta_m|\mu_1,\cdots, \mu_m\rangle$ is a symmetric polynomial in $\mu_1,\cdots,\mu_m$.
So are the expressions \eqref{A}, \eqref{B}, \eqref{C}, \eqref{D}. However, to arrive at the polynomial form
we have to cancel some singularities by factorizing the expressions in the right hand side. Factorisation
is a time consuming operation, it should be ultimately avoided. Actually, looking at the Slavnov formula one 
realises that it can be easily rewritten as a sum of Schur polynomials. Then with some effort one rewrites the right
hand sides of 
\eqref{A}, \eqref{B}, \eqref{C}, \eqref{D} as operators acting in the basis of Schur polynomials
(indexed by Young diagrams). In this way we completely eliminate time-consuming symbolic computations. 
In fact acting in the space of Young diagrams can be made very fast. 

\section{Final restriction of the number of variables}

Now for any given input data we construct the full Matsubara data which allow to compute, as explained
in the previous section, the quantities
$\langle\mathcal{O}\rangle_\mathrm{Md}$ and $\langle v_\al\rangle_\mathrm{Md}$ in the main formula \eqref{main}.

We start with $\mathcal{O}=\sigma_1^a\sigma_N^a$. It is interesting that in order to fix $C_\al$ up to $N=4$
we do not need to resort to non-trivial excited states $\Psi$ in the Matsubara data to fix the coefficients. Thus for $N=4$, absolutely trivial
computations give results allowing access to the correlation functions in any environment.
For $N=5$ we need one one-particle state $\Psi$. Starting from $N=8$ a few two-particle states are needed.
We always take more Matsubara data than the number of unknowns, so that the systems of equations to solve for $C_\al$ are
overdetermined. The compatibility of all these equations is an important check of the validity of our procedure. 

We computed up to $N=9$, and going further was  very hard. However,  analysing the  results we made the
following important observation. Decompose the operators  $v_\al$ in the fermionic basis to rewrite
\begin{align}\mathcal{O}\equiv 
\sum\limits_{I,J} D_{I,J}\  b^*_Ic^*_J\cdot\mathrm{I}\,.\nn
\end{align}
{\it A priori} $I,J$ are only lexicographically ordered, but we observe that the coefficients  $D_{I,J}$ vanish
unless
$$i_p\le j_p, \quad p=1,\cdots, \#(I)\,.$$
We shall write $I\preccurlyeq J$.
 
We want to reduce the number of unknowns using this observation.
The na\"ive idea consists in repeating the procedure of Section \ref{Basis} starting with
the subspace $\widetilde{\mathfrak{H}}^{(N)}_0$ spanned by $ b^*_Ic^*_J\cdot\mathrm{I}$ satisfying  \eqref{Cinv},
\eqref{lessN}
and 
$I\preccurlyeq J$.
However, the resulting space \eqref{resspace} would be too small, there would be no solutions to the equations
for $C_\al$. More careful consideration and further experiments show that the correct choice is
the subspace $\widetilde{\mathfrak{V}}^{(N)}$ of $ \widetilde{\mathfrak{H}}^{(N)}_0$ such that
$$\bigcap\limits_{i=N+1}^{2N+1} \mathrm{Im}\left(Q_{i}\big\vert_{\widetilde{\mathfrak{V}}^{(N)}}\right)\subset \mathfrak{N}\,.$$

This results in a drastic reduction of the dimension of the space. The dimensions up to $N=12$ are given in the table below.
\vskip .2cm

\scalebox{1}{\begin{tabular}{|r|r|r|r|r|r|r|r|r|r|r|r|}
  \hline
  $N$\ \ \  &2 & 3 & 4 &5&6&7&8&9&10&11&12\\
  \hline  
 \scalebox{.7}{$ \mathrm{dim}(\widetilde{\mathfrak{H}}^{(N)}_0)$}&1&3&11&26&99&253&1038&2816&12041&34062&148630 \\
\hline 
\ \scalebox{.7}{$\mathrm{dim}(\widetilde{\mathfrak{V}}^{(N)})$}&1&2&6&12&31&79&178&434&1141&2946&7888 \\
\hline
\end{tabular}}

\section{Results and conjectures}

We computed the OPE coefficients for $\sigma^a_1\sigma^a_N$ and $h_{1,2}h_{N-1,N}$ up to $N=11$.
Starting from $N=7$ the data becomes too large to be presented here, they are available upon request. 

With the OPE coefficients at hand we can compute the correlation functions for any Matsubara data, 
as well as the corresponding
matrix $\omega_{i,j}$ or equivalently the function $\omega(\la,\mu)$. In particular, for the
anti-ferromagnet at zero temperature and zero magnetic field we have
$$
\omega(\la,\mu) = -\half + 2 \log(2) + 
  \sum\limits_{k=1}^{\infty}(\la-\mu)^{2 k} \Bigl(2 \zeta(2 k + 1) (1 - 2^{-2 k}) - \half\Bigr)\,.$$
Let us present the numerical values for $G^{\sigma\sigma}$. Up to $N=8$ they are known from \cite{takahashi}, so,
we use the normalisation of this paper $G^{\sigma\sigma}(N)\to G^{\sigma\sigma}(N)/12$ to simplify the comparison.
\vskip .5cm
\centerline{
\begin{tabular}{|r|r|}
  \hline
  $N$ & $G^{\sigma\sigma}(N)/12\qquad\qquad\qquad $\\
\hline
  2 &-0.147715726853315103139077
\\
\hline
3 &0.0606797699564353014934941
\\
\hline
4 &-0.0502486272572352479593931
\\
\hline
5 &0.0346527769827281656556596
\\
\hline
6 &-0.0308903666476093257628751
\\
\hline
7 &0.0244467383279589065417695
\\
\hline
8 &-0.0224982227633722183770986
\\
\hline
9 &0.0189734169587321977494075
\\
\hline
10& -0.0177751064604679461357958
\\
\hline
11 &0.0155478493216075886422179
\\
\hline
\end{tabular}}

\vskip .5cm
\noindent
The last two values are already in a good agreement with the asymptotic formula of \cite{lukyanov}.

We can proceed with computations for finite temperature and other Matsubara environments, however,
the most interesting applications of our ``experimental" data consists in trying to guess general
structure of the OPE coefficients. We have some incomplete results in this direction which we expose now. 

Consider the decompositions
\begin{align}
&\sigma^1_1\sigma^a_N\equiv\sum_{I\preccurlyeq J}D_{I,J}^{\sigma\sigma}(N)\bb^*_I\cb^*_J\cdot\mathrm{I}\,,\label{decom}\\
&h_1h_{N-1}\equiv\sum_{I\preccurlyeq J}D_{I,J}^{hh}(N)\bb^*_I\cb^*_J\cdot\mathrm{I}\,.\nn
\end{align}
A first question which we would like to address is the behaviour of the coefficients as functions of $N$.
The experimental data show that they are polynomial in $N$ of degrees
\begin{align}
d^{\sigma\sigma}_{I,J}=
\half(|I|+|J|)+\#(I)-1\,,\qquad
d^{hh}_{I,J}=\half(|I|+|J|)+\#(I)-4\,.\label{degrees}\end{align}
Moreover, in these polynomials not all the coefficients are independent, we find that they are of the form
\begin{align}
D_{I,J}^{\sigma\sigma}(N)=\sum\limits_{s=l^{\sigma\sigma}_{I,J}}^{d^{\sigma\sigma}_{I,J}}X^{\sigma\sigma}_{s,I,J}\binom {N-2} s\,,\qquad
D_{I,J}^{hh}(N)=\sum\limits_{s=l^{hh}_{I,J}}^{d^{hh}_{I,J}}X^{hh}_{s,I,J}\binom {N-4} s\,.\label{coefX}
\end{align}
The lower limits $l^{\sigma\sigma}_{I,J}$, $l^{hh}_{I,J}$ are rather subtle. Lower bounds for them are
$$
l^{\sigma\sigma}_{I,J}\ge \mathrm{max}(J)-2
\,,\qquad l^{hh}_{I,J}\ge \mathrm{max}(J)-4
\,.
$$
These estimates show that the coefficients vanish if $\max(J)$ goes far from the rest of elements of $I\cup J$. 
However, if two elements of $J$ become large no vanishing happens. 
There are some additional simple reason for the coefficients to vanish which will be clear from examples below.
In certain cases we were even able to make general conjectures for the coefficients. Let us present them.

We begin with the simple case of $X^{\sigma\sigma}_{s,I,J}$ (coefficients in \eqref{coefX}) 
for the two-fermion case with $I=\{k-p\}$ and $J=\{k+p\}$.
In that case we have $3-p
\le 3$ terms in our formula. 
Since the space $\mathfrak{N}$ is empty for the zero fermion case 
we have for any $s$
\begin{align}\sum\limits _p X^{\sigma\sigma}_{s,\{k-p\},\{k+p\}}=0\,. \label{Qvan}
\end{align}
In particular, for $s=k-2$ there is only one term in this sum. Hence $ X^{\sigma\sigma}_{ k-2,\{k\},\{k\}}=0$, this is
the above mentioned additional reason for coefficients to vanish.
For nontrivial coefficients it is easy to guess from our data that
\begin{align}
& X^{\sigma\sigma}_{ k,\{k-p\},\{k+p\}}=2(-1)^{p-1}\(\binom{2}{ p}+2\binom{1}{ p}\)\,,\quad
 X^{\sigma\sigma}_{ k-1,\{k-p\},\{k+p\}}=2(-1)^{p-1}\binom{1}{ p}\,.\nn
\end{align}
Both coefficients vanish for $p>2$ in agreement with the previous explanation. 

It so happens that all the two fermions contribution to $G^{hh}(N)$  vanish.
The next simplest case is
that of four fermions contribution. 
For that case we use for the sets $I=\{k-p,i\}$, $J=\{k+p,j\}$ for $i+j$ even and $I=\{k-p,i\}$, $J=\{k+p+1,j\}$ for $i+j$ odd.
For fixed $i,j$ we shall vary $k$ and $p$ within the obvious limits.
Non-vanishing coefficients $X^{hh}$ for some $(k,p)$ correspond to boxes in a triangular array (see {\it Tab. 1} for $i=1,j=1$).
The vertical and horizontal sizes of the array are equal to $S=\left[\frac {i+j} 2\right]+3$.
Counting the rows by $p=0, \cdots , S-1$, and the columns by $r=0,\cdots, S-1$, then let
$s=r+k-4$ for $i+j$ even and $s=k+r-3$ for $i+j$ odd. For given $i$, $j$ we shall simplify notations: define 
$X_{r,p}(k):=X^{hh}_{s,\{k-p,1\},\{k+p,1\}}$ with $r$ and $s$ related as explained.
Sums over columns equal zero for the same reason as explained above, so, we do
not write down $X_{r,r}(k)$. Notice, in particular, that $X_{0,0}(k)=0$.

We start from the simplest case $i=j=1$. Non-vanishing coefficients correspond to boxes in the table\begin{align} 
&
\scalebox{1.5}{
\young(\bullet \ \ \  ,:\bullet \ \  ,::\bullet \  ,:::\bullet  )
}
{\it Tab.\,1}\nn
\end{align}

So, we shall not write formulae
for the boxes with bullets for the reasons which have been explained. 
The coefficients grow rather fast with $k$, so, at the first glance there is no hope to find a formula for them.
However, looking at them attentively, we observed that the simplest one (the second from the left in the first row)
is just
\begin{align}
X_{1,0}(k)=\frac 1 6 (3^{k-2}-1)\,.\nn
\end{align}
Based on this observation and the experimental data we guessed further
\begin{align}
&X_{2,0}(k)= \frac{5}8\cdot3^{k - 1} +\frac{ (2 k^2 + 4 k - 45)}{24}\,,\nn\\
&X_{3,0}(k)= \frac 5 4 \cdot 3^{k - 1} + \frac{2 k^2 + 8 k - 49}{12}\,,\nn\\
&X_{2,1}(k)=-\frac 5 2\cdot (3^{k - 2} - 1)\,,\nn\\
& X_{3,1}(k)=-\frac 5 8\cdot3^k - \frac{2 k^2 + 8 k - 159}{24}\,,\nn\\
&X_{3,2}(k)= \frac{27} 4\cdot (3^{k - 3} - 1) - \frac{(k - 3) (k + 7)} 6\,.\nn\end{align}
For certain reason $X_{3,2}(3)=0$. This is made explicit in the last line. We shall continue making
such vanishings explicit.

Let us consider now the case $i=1$, $j=2$. The table is the same as for the previous case.
From experimental data 
we guess the formulae
\begin{align}
&X_{1, 0}(k) := -\frac{k - 2}{24}\cdot (3^{k - 1} - 1) -\frac{(k - 1) (k + 2)}{12}\,,\nn\\
&X_{2,  
  0}(k)= -\frac{5 (k - 1)}{16} \cdot 3^{k - 1} - \frac{2 k^3 + 18 k^2 + 23 k + 21}{48}\,,\nn\\
  &X_{3, 0}(k) := -\frac{5 k}{32}\cdot3^k - \frac{6 k^3 + 52 k^2 + 89 k + 128}{96}\,,\nn\\
  &X_{2, 1}(k)= \frac{5 (k - 1)}{32}\cdot3^k + \frac{2 k^3 + 38 k^2 + 57 k + 63}{96}\,,\nn\\
  &X_{3, 1}(k):= \frac{k}{32}\cdot3^{k + 2} + \frac{2 k^3 + 20 k^2 + 31 k + 64}{32}\,,\nn\\
  &X_{3, 2}(k)= -\frac{5}{32} k\cdot3^k + \frac{2 k^3 + 4 k^2 + 23 k - 64}{96}\,.\nn
\end{align}

At this point one can think that the general formula consists of combinations with polynomial coefficients of $1$ and $3^k$. 
Actually, this is not the case: we were able to fix several additional coefficients observing that $6^k$ starts to appear. 

Consider $i=2,j=2$. In this case our data allow to define only three coefficients:
\begin{align}
X_{1,0}(k)&= \frac1 {125}\cdot (6^{k - 1} - 1)
  - \frac{(2 k + 1) }{144} \cdot(3^{k - 1} - 1)+ \frac{( k-1) ( 45 k-37 )}{1800}\nn\,,\\
  X_{2,0}(k)&= 
 \frac{49 }{250}\cdot(6^{k-1} - 1) + 
  \frac{ ( k^2- 19 k -41 )}{96} \cdot (3^{k - 1} - 1) \nn\\&+ \frac{( k-1) (
     25 k^2+ 210 k +179 )}{1200}\nn\,,\\
    X_{2,1}(k)&= -\frac{98}{375} \cdot(6^{k - 1}- 1)+ \frac{(6 k + 13) }{
   24}\cdot (3^{k - 1} - 1) - \frac{(45 k + 43) (k - 1)}{300}\nn\,.
\end{align}

Similarly for $i=1,j=3$:
\begin{align}
X_{1,0}(k)&=  -\frac{48 }{625}\cdot(6^{k - 2} - 1) + 
  \frac{  (7 k - 11)}{60} \cdot(3^{k - 2} - 1)- \frac{(k - 2) (5 k - 53)}{750} \nn\,,\\
  X_{2,0}(k&)=  -\frac{144\cdot49}{625} \cdot  (6^{k - 3} - 1)- 
  \frac{27 (2 k^2  -15 k- 18 )}{
    720}\cdot  (3^{k - 3} - 1) \nn\\&+ \frac{( k-3) ( 25 k^2- 465 k +10054 )}{3000}\nn\,,\\
    X_{2,1}(k)&=\frac{32\cdot49}{625}\cdot 6^{ k-2} - 2k\cdot 3^{ k-2}  + \frac{
 75 k^2 + 1105 k - 964}{1875}\nn\,.
\end{align}

Finally, we were able to guess one coefficient for each $i=2,j=3$ and \newline
$i=1,j=4$. They are
respectively
\begin{align}
X_{1,0}(k)&= 
-\frac{ 6(k-4)  }{625}\cdot(6^{k - 1} - 1) -\frac{(4 k^2-9k-8)}{
   240} \cdot (3^{k - 1} - 1 ) \nn\\& - \frac{( k-1) ( 65 k^2+ 91 k+96  )}{3000}\nn\,,\\
   X_{1,0}(k)&=
 \frac{12 ( k-4)}{625}\cdot  ( 6^{k - 1} - 1) +  \frac{ k^2+ 9 k-40  }{
   120} \cdot(3^{k - 1} - 1)\nn\\&+\frac{( k-1) ( 5 k^2+ 47 k+132  )} {500} \,.\nn
\end{align}

An important question is that of the general structure. Is it true that the next exponent to appear will be $9^k$?
This should be possible to answer considering the cases $i+j=6$ for which we observe experimentally faster
growth than $6^k$. Unfortunately, our experimental data do not allow to guess a general formula for
this case.

For four fermions contributions to $G^{\sigma\sigma}$ the coefficients are more complicated, we shall not
present them because we know only very few. We would mention, however, that the structure looks similar,
but $6^k$ starts to appear from the very beginning.

\section{Conclusion}

We have demonstrated the power of the method to fix the OPE coefficients
for the XXX model based on the fermionic basis. Somebody with better computer skills than ours
may probably produce some more data. For example $N=12$ should be reachable. 

The main problem which we could not solve for the moment is finding
general formulae for the coefficients even for the four fermion contributions to $G^{hh}$. For example,
the coefficient $X_{r,p}(k)$ is known for $\{i,j\}=\{1,1\}, \{1,2\},\{1,3\},\{1,4\},\{2,2\},\{2,3\},$ but still
we were unable to find a conjecture for general $i,j$. Probably, an independent look is needed?

 \section*{Acknowledgements}

PDF is partially supported by the Morris and Gertrude Fine endowment. FS would like to thank
H. Boos, M. Jimbo and T. Miwa for stimulating discussions.





\begin{thebibliography}{9}




\bibitem{HGSV}
 M.~Jimbo, T.~Miwa, F. Smirnov,
\newblock Hidden {Grassmann} structure in the {XXZ} model {V}: {sine-Gordon}
 model.
\newblock {\em Lett. Math. Phys.}, {\bf 96} (2011) 325--365
\bibitem{NS} S.~Negro, F.~Smirnov
\newblock
On one-point functions for sinh-Gordon model at finite temperature
\newblock{\em Nuclear Physics} {\bf B 875(FS)} (2013) 166-185 

\bibitem{FST}  L.D.~Faddeev, E.K.~Sklyanin, L.A.~takhtajan
\newblock
The quantum inverse problem method {\it  Theoretical and Mathematical Physics} {\bf 40} (1980) 688-711

\bibitem{BIK} V.E.~Korepin, N.M.~Bogoliubov, A.G.~Izergin \newblock Quantum inverse scattering method and
correlation functions {\it Cambridge University Press} (1993) 

\bibitem{Slavnov} N.A.~Slavnov
\newblock
Calculation of scalar products of wave functions and form factors in the framework of the alcebraic Bethe ansatz
{\it  Theoretical and Mathematical Physics} {\bf 79} (1989) 502-508

\bibitem{HGSIII}
M.~Jimbo, T.~Miwa, and F.~Smirnov.
\newblock Hidden {Grassmann} structure in the {XXZ} model {III}: {Introducing
  Matsubara} direction.
\newblock {\em J. Phys. A} {\bf 42} (2009)  304018 (31pp)

\bibitem{HGSII}
H.~Boos, M.~Jimbo, T.~Miwa, F.~Smirnov, and Y.~Takeyama.
\newblock Hidden {Grassmann} structure in the {XXZ} model {II : Creation}
  operators.
\newblock {\em Commun. Math. Phys.} {\bf 286} (2009) 875--932. 

\bibitem{completness}
H.~Boos, M.~Jimbo, T.~Miwa, F.~Smirnov.\newblock
\newblock Completeness of a fermionic basis in the homogeneous {$XXZ$} model.
\newblock {\em J. Math. Phys.}, {\bf 50} (2009) 095206 

\bibitem{FB} 
H.~Boos, M.~Jimbo, T.~Miwa, F.~Smirnov, and Y.~Takeyama.
\newblock 
Fermionic basis for space of operators in the XXZ model.
\newblock 
{\it SISSA Proceedings of Science} (2007), Paper 015, 34

\bibitem{BGKS}
H.~Boos, F.~G{\"o}hmann, A.~Kl{\"u}mper and J.~Suzuki,  
\newblock
Factorization of the finite temperature correlation functions of 
the XXZ chain in a magnetic field,  
\newblock {\em J. Phys. A} \textbf{40} (2007) 10699--10727. 

\bibitem{takahashi} J.~Sato, M.~ Shiroishi, M.~Takahashi
\newblock
Exact evaluation of density matrix elements for the Heisenberg chain
{\it J.Stat.Mech.} {\bf 0612} (2006) P12017
 \bibitem{lukyanov}
 S.~ Lukyanov, V.~ Terras.
 \newblock
 Long-distance asymptotics of spinÐspin correlation functions for the XXZ spin chain
{\it Nuclear Physics} {\bf  B 654} (2003), 323-356

\end{thebibliography}
 \end{document}